\documentclass[twocolumn,twoside,preprintnumbers,amsmath,amssymb,pacs]{revtex4}
\usepackage{epsfig}
\usepackage{graphicx}
\usepackage{dcolumn}
\usepackage{bm}

\usepackage{fancyhdr}

\usepackage{pslatex}

\newcommand{\pslash}{p \! \! \! /}

\newcommand{\Nf}{N_{\!f}}

\newcommand{\MSbar}{\overline{\mbox{MS}}}
\newcommand{\oB}{\overline{B}}
\newcommand{\occ}{\overline{c}}
\newcommand{\oG}{\overline{G}}
\newcommand{\p}{\partial}
\newcommand{\NA}{N_{\!A}}

\begin{document}

\title{{\Large A local non-Abelian gauge invariant action stemming from the nonlocal operator $F_{\mu\nu}(D^2)^{-1}F_{\mu\nu}$  }}
\author{D. Dudal$^a$}\email{david.dudal@ugent.be}\altaffiliation{Postdoctoral fellow of the
\emph{Special Research Fund} of Ghent University.}
\author{M.A.L. Capri$^b$}
\email{marcio@dft.if.uerj.br}
\author{J.A. Gracey$^{c}$}
    \email{jag@amtp.liv.ac.uk}
\author{V.E.R. Lemes$^{b}$}\email{vitor@dft.if.uerj.br}
\author{R.F. Sobreiro$^b$}
 \email{sobreiro@dft.if.uerj.br}
\author{S.P. Sorella$^b$}
\email{sorella@uerj.br} \altaffiliation{Work supported by FAPERJ,
Funda{\c c}{\~a}o de Amparo {\`a} Pesquisa do Estado do Rio de
Janeiro, under the program {\it Cientista do Nosso Estado},
E-26/151.947/2004.}
\author{H. Verschelde$^a$}
 \email{henri.verschelde@ugent.be}
 \affiliation{\vskip 0.1cm $^a$
Ghent University
\\ Department of Mathematical
Physics and Astronomy \\ Krijgslaan 281-S9 \\ B-9000 Gent,
Belgium\\\\\vskip 0.1cm $^b$
 UERJ - Universidade do Estado do Rio de
Janeiro\\Rua S\~{a}o Francisco Xavier 524, 20550-013
Maracan\~{a}\\Rio de Janeiro, Brasil\\\\
\vskip 0.1cm $^c$ Theoretical Physics Division\\ Department of
Mathematical Sciences\\ University of Liverpool\\ P.O. Box 147,
Liverpool, L69 3BX, United Kingdom }


\begin{abstract}
We report on the nonlocal gauge invariant operator of dimension two,
$F_{\mu\nu} (D^2)^{-1} F_{\mu\nu}$. We are able to localize this
operator by introducing a suitable set of (anti)commuting
antisymmetric tensor fields.  Starting from this, we succeed in
constructing a local gauge invariant action containing a mass
parameter, and we prove the renormalizability to all orders of
perturbation theory of this action in the linear covariant gauges
using the algebraic renormalization technique. We point out the
existence of a nilpotent BRST symmetry. Despite the additional
(anti)commuting tensor fields and coupling constants, we prove that
our model in the limit of vanishing mass is equivalent with ordinary
massless Yang-Mills theories by making use of an extra symmetry in
the massless case. We also present explicit renormalization group
functions at two loop order in the $\MSbar$ scheme.
\end{abstract}

\maketitle

\thispagestyle{fancy} \setcounter{page}{1}

\section{Introduction}
We shall consider pure Euclidean $SU(N)$ Yang-Mills theories with
action
\begin{equation}\label{1}
{ S_{YM}}=\frac{1}{4}\int d^4 x\;F_{\mu \nu }^{a}F^{a}_{\mu \nu }\;,
\end{equation}
where $A_{\mu }^{a}$, $a=1,...,N^{2}-1$ is the gauge boson field,
with associated field strength
\begin{equation}\label{3}
F^{a}_{\mu \nu }=\partial _{\mu }A_{\nu }^{a}-\partial _{\nu }A_{\mu
}^{a}+gf^{abc}A_{\mu }^{b}A_{\nu }^{c}\;.
\end{equation}
The theory (\ref{1}) is invariant with respect to the local gauge
transformations
\begin{equation}\label{4}
    \delta A_\mu^a = D_{\mu}^{ab}\omega^b\;,
\end{equation}
with
\begin{equation}\label{5}
    D_{\mu}^{ab}=\partial_\mu\delta^{ab}-gf^{abc}A_\mu^c\;,
\end{equation}
denoting the adjoint covariant derivative.

As it is well known, the theory (\ref{1}) is asymptotically free
\cite{Gross:1973id, Politzer:1973fx}, i.e. the coupling becomes
smaller at higher energies and vice versa. At very high energies,
the interaction is weak and the gluons can be considered as almost
free particles. However, in spite of the progress in the last
decades, we still lack a satisfactory understanding of the behaviour
of Yang-Mills theories in the low energy regime. Here the coupling
constant of the theory is large and nonperturbative effects have to
be taken into account.

The introduction of condensates, i.e. the (integrated) vacuum
expectation value of certain operators, allows one to parametrize
certain nonperturbative effects arising from the infrared sector of
e.g. the theory described by (\ref{1}). Via the Operator Product
Expansion (OPE) (viz. short distance expansion), which is applicable
to local operators, one can relate these condensates to power
corrections which give nonperturbative information in addition to
the perturbatively calculable results. If one wants to consider the
possible effects of condensates on physical quantities in a gauge
theory, quite clearly only gauge invariant operators should be
considered. The most famous example is the dimension 4 gluon
condensate $\left\langle\alpha_s F_{\mu\nu}^2\right\rangle$, giving
rise to $\frac{1}{Q^4}$ power corrections. Via the SVZ
(Shifman-Vainshtein-Zakharov) sum rules \cite{Shifman:1978bx}, one
can extract phenomenological estimates for $\left\langle\alpha_s
F_{\mu\nu}^2\right\rangle$.

In recent years, a great deal of interest arose in dimension 2
condensates in gauge theories. Most attention was paid to the gluon
condensate $\left\langle A_\mu^2\right\rangle$ in the Landau gauge,
due to the work of \cite{Gubarev:2000nz,Gubarev:2000eu}, as the
quantity
\begin{equation}\label{intro1}
    \left\langle A^2_{\min}\right\rangle\equiv\min_{U\in SU(N)}\frac{1}{VT}\int
    d^4x \left\langle\left(A_\mu^U\right)^2\right\rangle\;,
\end{equation}
which is gauge invariant due to the minimization along the gauge
orbits, could be physically relevant. In fact, as shown in
\cite{Gubarev:2000nz,Gubarev:2000eu} in the case of compact QED, the
quantity $\left\langle A^2_{\min}\right\rangle$ seems to be useful
in order to detect the presence of nontrivial field configurations
like monopoles. One can show that $A_{\min}^2$ can be written as an
infinite series of nonlocal terms, see
\cite{Lavelle:1995ty,Capri:2005dy} and references therein,
namely\begin{widetext}
\begin{eqnarray}
A_{\min }^{2} &=&\int d^{4}x\left[ A_{\mu }^{a}\left( \delta _{\mu \nu }-\frac{%
\partial _{\mu }\partial _{\nu }}{\partial ^{2}}\right) A_{\nu
}^{a}-gf^{abc}\left( \frac{\partial _{\nu }}{\partial ^{2}}\partial
A^{a}\right) \left( \frac{1}{\partial ^{2}}\partial {A}^{b}\right)
A_{\nu }^{c}\right] \;+O(A^{4})\;.  \label{intro2}
\end{eqnarray}
\end{widetext}
Since the operator $A_{\min}^2$ is nonlocal, it falls beyond the
applicability of the OPE annex sum rules, which refer to local
operators.

However, in the Landau gauge, $\p_\mu A_\mu=0$, all nonlocal terms
of expression (\ref{intro2}) drop out, so that $A_{\min }^{2}$
reduces to the local operator $A_\mu^2$, hence the interest in the
Landau gauge and its dimension two gluon condensate $\left\langle
A_\mu^2\right\rangle$. A complication is that the explicit
determination of the absolute minimum of $A_\mu^2$ along its gauge
orbit, and moreover of its vacuum expectation value, is a very
delicate issue intimately related to the problem of Gribov copies
\cite{Gribov:1977wm,Semenov,Zwanziger:1990tn,Dell'Antonio:1989jn,Dell'Antonio:1991xt,vanBaal:1991zw}.

Nevertheless, some nontrivial results were proven concerning the
operator $A_\mu^2$. In particular, we mention its multiplicative
renormalizability to all orders of perturbation theory, in addition
to an interesting and numerically verified relation concerning its
anomalous dimension \cite{Dudal:2002pq,Gracey:2002yt}. An effective
potential approach consistent with renormalizability and
renormalization group requirements for local composite operators
(LCO) has also been worked out for this operator, giving further
evidence of a nonvanishing condensate $\left\langle
A_\mu^2\right\rangle\neq0$, which  lowers the nonperturbative vacuum
energy \cite{Verschelde:2001ia}. The LCO method yields an effective
gluon mass squared $m_g^2\sim\left\langle A_\mu^2\right\rangle$ of a
few hundred MeV
\cite{Verschelde:2001ia,Browne:2004mk,Gracey:2004bk,Browne:2003uv}.

In \cite{Lavelle:1988eg}, it was already argued that gauge
(in)variant condensates could also influence gauge variant
quantities such as the gluon propagator. An OPE argument based on
lattice simulations in the Landau gauge has indeed provided evidence
that a condensate $\left\langle A_\mu^2\right\rangle$ could account
for quadratic power corrections of the form $\sim\frac{1}{Q^2}$,
reported in the running of the coupling constant as well as in the
gluon propagator, see e.g.
\cite{Kondo:2001nq,Boucaud:2001st,Boucaud:2002nc,RuizArriola:2004en,Furui:2005bu,Furui:12ks}.
This OPE approach allows one to obtain an estimate for the soft part
$\left\langle A_\mu^2\right\rangle_{IR}$ originating from the
infrared sector. The OPE can also be employed to relate this
condensate to an effective gluon mass \cite{Kondo:2001nq}.

The presence of mass parameters in the gluon propagator have also
been advocated from the lattice perspective several times
\cite{Marenzoni:1994ap,Leinweber:1998uu,Langfeld:2001cz,Amemiya:1998jz,Bornyakov:2003ee,Cucchieri:2004mf},
whilst effective gluon masses also found phenomenological use
\cite{Parisi:1980jy,Field:2001iu}.

A somewhat weak point about the operator $A^2_{\min}$ is that it is
unclear how to deal with it in gauges other than the Landau gauge.
Till now, it seems hopeless to prove its renormalizability out of
the Landau gauge. In fact, at the classical level, adding
(\ref{intro1}) to the Yang-Mills action is equivalent to add the
so-called Stueckelberg action, which is known to be not
renormalizable \cite{Ruegg:2003ps,vanDam:1970vg}. We refer to
\cite{Capri:2005dy} for details and references. As already
mentioned, also the OPE becomes useless outside the Landau gauge for
this particular operator.

In other gauges, there can be found other renormalizable local
operators, which condense and give rise to a dynamical gluon mass.
Next to the Landau gauge
\cite{Verschelde:2001ia,Dudal:2002pq,Browne:2004mk,Gracey:2004bk,Browne:2003uv,Sobreiro:2004us,Dudal:2005na}
the maximal Abelian \cite{Dudal:2004rx,Capri:2005tj}, linear
covariant \cite{Dudal:2003np,Dudal:2003by,Sobreiro:2005vn} and
Curci-Ferrari gauges \cite{Dudal:2003gu,Dudal:2003pe} have been
investigated in the past.

The relevant operators in these other gauges are however gauge
variant, and as a consequence also the effective gluon mass. From
this perspective, it is worthwhile to find out whether a gauge
invariant framework might be found for a dynamical mass, and related
to it for $\frac{1}{Q^2}$ power corrections.

In order to have a starting point, we need a dimension 2 operator
that is gauge invariant. This necessarily implies a nonlocal
operator, since gauge invariant local operators of dimension 2 do
not exist. We would also need a consistent calculational framework,
which requires an action only containing local terms. Therefore, we
should find an operator that can be localized by means of a finite
set of auxiliary fields, in such a way that the local gauge
invariance is respected. As $A_{\min }^{2}$ looks a bit hopeless
from this viewpoint as it is a infinite series of nonlocal terms
(\ref{intro2}), we moved our attention instead to the nonlocal gauge
invariant operator
\begin{equation}\label{op}
\mathcal{O}\equiv\frac{1}{VT}\int d^{4}xF_{\mu \nu }^{a}\left[
\left( D^{2}\right) ^{-1}\right] ^{ab}F_{\mu \nu }^{b}\;.
\end{equation}
This operator caught already some attention in 3 dimensional gauge
theories in relation to a dynamical mass generation
\cite{Jackiw:1997jg}.

In the following sections, we shall show that the operator
(\ref{op}) can be localized, giving rise to a local, classically
gauge invariant action. Afterwards, we discuss how to investigate
the renormalizability of the action once quantized. Eventually, we
need to introduce a slightly more general classical action in order
to obtain a quantum action that is renormalizable to all orders of
perturbation theory. In the case of vanishing mass, the equivalence
of our action with usual Yang-Mills theories can be shown. We shall
point out the existence of a naturally extended version of the usual
BRST symmetry. Before turning to conclusions, we explicitly give
various renormalization group functions, verifying the
renormalizability at the practical level.

\section{Construction of the action and its renormalizability at the quantum level}
\subsection{The action at the classical level}
We can add the operator (\ref{op}) to the Yang-Mills action as a
mass term via
\begin{equation}
S_{YM}+S_{\mathcal{O}}\;,  \label{ymop}
\end{equation}
with
\begin{equation}
S_{\mathcal{O}}=-\frac{m^{2}}{4}\int d^{4}xF_{\mu \nu }^{a}\left[
\left( D^{2}\right) ^{-1}\right] ^{ab}F_{\mu \nu }^{b}\;.
\label{massop}
\end{equation}
As we have discussed in \cite{Capri:2005dy}, the action (\ref{ymop})
can be localized by introducing a pair of complex bosonic
antisymmetric tensor fields, $\left( B_{\mu \nu
}^{a},\overline{B}_{\mu \nu }^{a}\right) $, and a pair of complex
anticommuting antisymmetric tensor fields, $\left( \overline{G}_{\mu
\nu }^{a},G_{\mu \nu }^{a}\right) $, belonging to the adjoint
representation, according to which
\begin{eqnarray}
e^{-S_{\mathcal{O}}}&=&\int D\overline{B}DBD\oG DG\exp \left[
-\left( \frac{1}{4}\int d^{4}x\overline{B}_{\mu \nu }^{a}D_{\sigma
}^{ab}D_{\sigma }^{bc}B_{\mu \nu }^{c}\right.\right.\nonumber\\&-&\left.\left.\frac{1}{4}\int {%
d^{4}x}\overline{G}_{\mu \nu }^{a}D_{\sigma }^{ab}D_{\sigma
}^{bc}G_{\mu
\nu }^{c}+\frac{im}{4}\int d^{4}x\left( B-\overline{B%
}\right) _{\mu \nu }^{a}F_{\mu \nu }^{a}\right) \right] \nonumber\\
\label{loc2}
\end{eqnarray}
Therefore, we obtain a classical local action which reads
\begin{equation}
S_{YM}+S_{BG}+S_{m}\;,  \label{action1}
\end{equation}
where
\begin{eqnarray}
S_{BG} &=&\frac{1}{4}\int d^{4}x\left( \overline{B}_{\mu \nu
}^{a}D_{\sigma }^{ab}D_{\sigma }^{bc}B_{\mu \nu
}^{c}-\overline{G}_{\mu \nu }^{a}D_{\sigma
}^{ab}D_{\sigma }^{bc}G_{\mu \nu }^{c}\right) \;,  \nonumber \\
S_{m} &=&\frac{im}{4}\int d^{4}x\left( B-\overline{B}\right) _{\mu
\nu }^{a}F_{\mu \nu }^{a}\;,  \label{actions2}
\end{eqnarray}
which is left invariant by the gauge transformations
\begin{eqnarray}
\delta A_{\mu }^{a} &=&-D_{\mu }^{ab}\omega ^{b}\;,  \nonumber \\
\delta B_{\mu \nu }^{a} &=&gf^{abc}\omega ^{b}B_{\mu \nu
}^{c}\;,\;\; \delta \overline{B}_{\mu \nu }^{a} =gf^{abc}\omega
^{b}\overline{B}_{\mu \nu }^{c}\;,
\nonumber \\
\delta G_{\mu \nu }^{a} &=&gf^{abc}\omega ^{b}G_{\mu \nu
}^{c}\;,\;\; \delta \overline{G}_{\mu \nu }^{a} =gf^{abc}\omega
^{b}\overline{G}_{\mu \nu }^{c}\;. \label{gtm}
\end{eqnarray}
\subsection{The action at the quantum level}
In order to discuss the renormalizability of (\ref{action1}), we
relied on a method introduced by Zwanziger in
\cite{Zwanziger:1989mf,Zwanziger:1992qr}. Instead of using
(\ref{action1}) with $m$ coupled to the composite operators
$B_{\mu\nu}^aF_{\mu\nu}^a$ and $B_{\mu\nu}^aF_{\mu\nu}^a$, we
introduce 2 suitable external sources $V_{\rho\sigma\mu\nu}$ and
$\overline{V}_{\rho\sigma\mu\nu}$ and replace $S_m$ by
\begin{equation}
\frac{1}{4}\int d^{4}x\left( V_{\sigma \rho \mu \nu
}\overline{B}_{\sigma \rho }^{a}F_{\mu \nu
}^{a}-\overline{V}_{\sigma \rho \mu \nu }B_{\sigma \rho }^{a}F_{\mu
\nu }^{a}\right) \;. \label{rs}
\end{equation}
At the end, the sources $V_{\sigma \rho \mu \nu }(x)$,
$\overline{V}_{\sigma \rho \mu \nu }(x)$ are required to attain
their physical value, namely
\begin{equation}
\overline{V}_{\sigma \rho \mu \nu }\Big|_{\mathrm{phys}}=V_{\sigma \rho \mu \nu }%
\Big|_{\mathrm{phys}}=-\frac{im}{2}\left( \delta _{\sigma \mu
}\delta _{\rho \nu }-\delta _{\sigma \nu }\delta _{\rho \mu }\right)
\;, \label{ps}
\end{equation}
so that (\ref{rs}) reduces to $S_m$ in the physical limit.

From now on, we assume the linear covariant gauge fixing,
implemented through
\begin{equation}
S_{gf}=\int d^{4}x\left( \frac{\alpha }{2}b^{a}b^{a}+b^{a}\partial
_{\mu }A_{\mu }^{a}+\overline{c}^{a}\partial _{\mu }D_{\mu
}^{ab}c^{b}\right) \;, \label{lg}
\end{equation}

In \cite{Capri:2005dy}, we wrote down a list of symmetries enjoyed
by the action
\begin{equation}
S_{YM}+S_{BG}+S_{gf}\;,
\end{equation}
i.e. in absence of the sources. Let us only mention here the BRST
symmetry, generated by the nilpotent transformation $s$ given by
\begin{eqnarray}
sA_{\mu }^{a} &=&-D_{\mu }^{ab}c^{b}\;,\;\;  sc^{a} =\frac{g}{2}f^{abc}c^{a}c^{b}\;,  \nonumber \\
sB_{\mu \nu }^{a} &=&gf^{abc}c^{b}B_{\mu \nu }^{c}+G_{\mu \nu
}^{a}\;,\;\; s\overline{B}_{\mu \nu }^{a}
=gf^{abc}c^{b}\overline{B}_{\mu \nu }^{c}\;,
\nonumber\\
sG_{\mu \nu }^{a} &=&gf^{abc}c^{b}G_{\mu \nu }^{c}\;,\;\;
s\overline{G}_{\mu \nu }^{a} =gf^{abc}c^{b}\overline{G}_{\mu \nu
}^{c}+\overline{B}_{\mu
\nu }^{a}\;,  \nonumber \\
s\overline{c}^{a} &=&b^{a} \;,\;\; sb^{a} =0\;,\;\;s^{2} =0\;.
\label{bi}
\end{eqnarray}
It turns out that one can introduce all the necessary external
sources in a way consistent with the starting symmetries. This
allows to write down several Ward identities by which the most
general counterterm is restricted using the algebraic
renormalization formalism \cite{Capri:2005dy,Piguet:1995er}. After a
very cumbersome analysis, it turns out that the action
(\ref{action1}) must be modified to
\begin{eqnarray}
  S_{phys} &=& S_{cl} +S_{gf}\;,\label{completeaction}
  \end{eqnarray}
  with
\begin{widetext}
  \begin{eqnarray}
  S_{cl}&=&\int d^4x\left[\frac{1}{4}F_{\mu \nu }^{a}F_{\mu \nu }^{a}+\frac{im}{4}(B-\overline{B})_{\mu\nu}^aF_{\mu\nu}^a
  +\frac{1}{4}\left( \overline{B}_{\mu \nu
}^{a}D_{\sigma }^{ab}D_{\sigma }^{bc}B_{\mu \nu
}^{c}-\overline{G}_{\mu \nu }^{a}D_{\sigma }^{ab}D_{\sigma
}^{bc}G_{\mu \nu
}^{c}\right)\right.\nonumber\\
&-&\left.\frac{3}{8}%
m^{2}\lambda _{1}\left( \overline{B}_{\mu \nu }^{a}B_{\mu \nu
}^{a}-\overline{G}_{\mu \nu }^{a}G_{\mu \nu }^{a}\right)
+m^{2}\frac{\lambda _{3}}{32}\left( \overline{B}_{\mu \nu
}^{a}-B_{\mu \nu }^{a}\right) ^{2}+
\frac{\lambda^{abcd}}{16}\left( \overline{B}_{\mu\nu}^{a}B_{\mu\nu}^{b}-\overline{G}_{\mu\nu}^{a}G_{\mu\nu}^{b}%
\right)\left( \overline{B}_{\rho\sigma}^{c}B_{\rho\sigma}^{d}-\overline{G}_{\rho\sigma}^{c}G_{\rho\sigma}^{d}%
\right) \right]\;, \label{completeactionb}\label{lcg}
\end{eqnarray}
\end{widetext}
in order to have renormalizability to all orders of perturbation
theory. We notice that we had to introduce a new invariant quartic
tensor coupling $\lambda^{abcd}$, subject to the generalized Jacobi
identity
\begin{equation}\label{jacobigen}
    f^{man}\lambda^{mbcd}+f^{mbn}\lambda^{amcd}+f^{mcn}\lambda^{abmd}+f^{mdn}\lambda^{abcm}=0\,,
\end{equation}
and the symmetry constraints
\begin{eqnarray}
\lambda^{abcd}=\lambda^{cdab} \;, \nonumber \\
\lambda^{abcd}=\lambda^{bacd} \;, \label{abcd}
\end{eqnarray}
as well as two new mass couplings $\lambda_1$ and $\lambda_3$.
Without the new couplings, i.e. when $\lambda_1\equiv0$,
$\lambda_3\equiv0$, $\lambda^{abcd}\equiv0$, the previous action
would not be renormalizable. We refer to
\cite{Capri:2005dy,Capri:2006ne} for all the details. We also notice
that the novel fields $B_{\mu\nu}^a$, $\overline{B}_{\mu\nu}^a$,
$G_{\mu\nu}^a$ and $\overline{G}_{\mu\nu}^a$  are no longer
appearing at most quadratically. As it should be expected, the
classical action $S_{cl}$ is still gauge invariant w.r.t.
(\ref{gtm}).
\section{Further properties of the action}
\subsection{Existence of a BRST symmetry with a nilpotent charge}
The BRST transformation (\ref{bi}) no longer generates a symmetry of
the action $S_{phys}$. However, we are able to define a natural
generalization of the usual BRST symmetry that does constitute an
invariance of the gauge fixed action (\ref{completeaction}). Indeed,
after inspection, one shall find that
\begin{eqnarray}
  \widetilde{s}S_{phys} &=& 0\;, \nonumber\\
  \widetilde{s}^2 &=& 0\;,
\end{eqnarray}
with
\begin{eqnarray}
\widetilde{s} A_{\mu }^{a} &=&-D_{\mu }^{ab}c ^{b}\;,\;\;  \widetilde{s} c^{a} =\frac{g}{2}f^{abc}c^ac ^{b}\;,  \nonumber \\
 \widetilde{s} B_{\mu \nu }^{a} &=&gf^{abc}c ^{b}B_{\mu \nu }^{c}\;,\;\;\widetilde{s} \overline{B}_{\mu \nu }^{a} =gf^{abc}c
^{b}\overline{B}_{\mu \nu }^{c}\;,
\nonumber \\
\widetilde{s} G_{\mu \nu }^{a} &=&gf^{abc}c ^{b}G_{\mu \nu
}^{c}\;,\;\;\widetilde{s} \overline{G}_{\mu \nu }^{a} =gf^{abc}c
^{b}\overline{G}_{\mu
\nu }^{c}\;,\nonumber\\
\widetilde{s}\overline{c}^{a} &=&b^a\;,\;\; \widetilde{s} b^{a}
=0\;.
 \label{brst3}
\end{eqnarray}
Hence, the action $S_{phys}$ is invariant with respect to a
nilpotent BRST transformation $\widetilde{s}$. We obtained thus a
gauge field theory, described by the action $S_{phys}$,
(\ref{completeaction}), containing a mass term, and which has the
property of being renormalizable, while nevertheless a nilpotent
BRST transformation expressing the gauge invariance after gauge
fixing exists simultaneously. It is clear that $\widetilde{s}$
stands for the usual BRST transformation, well known from
literature, on the original Yang-Mills fields, whereas the gauge
fixing part $S_{gf}$ given in (\ref{lcg}) can be written as a
$\widetilde{s}$-variation, ensuring that the gauge invariant
physical operators shall not depend on the choice of the gauge
parameter \cite{Piguet:1995er}.

\subsection{Existence of a ``supersymmetry'' when $m\equiv0$}
We define a nilpotent (anti-commuting) transformation $\delta_s$ as
\begin{eqnarray}\label{ss}
\delta_s B_{\mu\nu}^a &=& G_{\mu\nu}^a \;,\qquad \delta_s G_{\mu\nu}^a =0 \;,\nonumber\\
\delta_s \overline{G}_{\mu\nu}^a &=& \overline{B}_{\mu\nu}^a \;,
\qquad \delta_s \overline{B}_{\mu\nu}^a = 0 \;,\nonumber\\ \delta_s
(\textrm{rest})&=&0 \;.
\end{eqnarray}
Then one easily verifies that (\ref{ss}) generates a
``supersymmetry'' of the action $S_{phys}^{m\equiv0}$ since
\begin{eqnarray}
    \delta_s S_{phys}^{m\equiv0}&=&0\;,
\end{eqnarray}
with
\begin{eqnarray}\label{ss2}
    \delta_s^2&=&0\;.
    \end{eqnarray}
Taking another look at the transformations $s$ and $\widetilde{s}$,
respectively given by (\ref{bi}) and (\ref{brst3}), one recognizes
that
\begin{eqnarray}
  s &=& \widetilde{s}+\delta_s\;, \nonumber\\
  \{\delta_s,\widetilde{s}\} &=& 0\;.
\end{eqnarray}
Since $\delta_s$ is a nilpotent operator, it possesses its own
cohomology, which is easily identified with polynomials in the
original Yang-Mills fields $\{A^a_{\mu}, b^a, c^a, {\overline c}^a
\}$. The auxiliary tensor fields, $\{B_{\mu\nu}^a,{\overline
B}_{\mu\nu}^a, G_{\mu\nu}^a, {\overline G}_{\mu\nu}^a\}$, do not
belong to the cohomology of $\delta_s$, because they form pairs of
doublets \cite{Piguet:1995er}. This fact can be brought to use in
the following subsection.

\subsection{Equivalence with Yang-Mills gauge theory when
$m\equiv0$} If the mass $m\equiv0$, we would expect that the action
(\ref{completeaction}) would be equivalent with the usual Yang-Mills
gauge theory, since in the nonlocal formulation (\ref{massop}), we
would have introduced ``nothing''. In the local renormalizable
formulation (\ref{completeaction}), this would also be trivially
true when $\lambda^{abcd}\equiv0$ as then we would only have added a
-although quite complicated- unity to the Yang-Mills action.
Unfortunately, since renormalization forbids setting
$\lambda^{abcd}=0$, we must find another argument to relate
$S_{phys}^{m\equiv0}$ to the usual Yang-Mills gauge theory. As
proven in \cite{Capri:2006ne}, the ``supersymmetry'' $\delta_s$ of
(\ref{ss}) can be used to show that
\begin{equation}\label{hom1}
    \left\langle G_n(x_1,\ldots,x_n)\right\rangle_{S_{YM}+S_{gf}}\equiv \left\langle
    G_n(x_1,\ldots,x_n)\right\rangle_{S_{phys}^{m\equiv0}}\;,
\end{equation}
where
\begin{widetext}
\begin{eqnarray}\label{dd3}
G_n(x_1,\ldots,x_n)&=& A(x_1)\ldots
A(x_i)\occ(x_{i+1})\ldots\occ(x_{j})c(x_{j+1})\ldots
c(x_{k})b(x_{k+1})\ldots
    b(x_n)\;,
\end{eqnarray}
\end{widetext}
is a generic Yang-Mills functional. The expectation value of any
Yang-Mills Green function, constructed from the fields
$\left\{A_\mu^a,c^a,\occ^a,b^a\right\}$ and calculated with the
original (gauge fixed) Yang-Mills action $S_{YM}+S_{gf}$, is thus
identical to the one calculated with the massless action
$S_{phys}^{m\equiv0}$, where it is of course assumed that the gauge
freedom of both actions has been fixed by an identical gauge fixing.

The foregoing result also reflects on the renormalization group
functions. As usual, we employ a massless renormalization scheme
known as the $\MSbar$ scheme. As a consequence, we can set $m=0$ in
order to extract the ultraviolet behaviour. Using (\ref{hom1}), we
conclude that all the renormalization group functions of the
original Yang-Mills quantities are not affected by the presence of
the extra fields or couplings. This fact shall be explicitly
verified in the next section.
\section{Explicit renormalization at two loop order}
Having proven the renormalizability of the action
(\ref{completeaction}), we shall now compute explicitly the two loop
anomalous dimension of the fields and the one loop $\beta$-function
of the tensor coupling $\lambda^{abcd}$. The corresponding details
can be found in \cite{Capri:2005dy} for one loop results, while two
loop results are discussed in \cite{Capri:2006ne}.

We have regarded the mass operator as an insertion and split the
Lagrangian into a free piece involving massless fields with the
remainder being transported to the interaction Lagrangian. To
renormalize the operator, we insert it into a massless Green
function, after the fields and couplings have been renormalized in
the massless Lagrangian. An attractive feature of the massless field
approach is that we can use the {\sc Mincer} algorithm to perform
the actual computations. This algorithm, \cite{Gorishnii:1989gt},
written in the symbolic manipulation language {\sc Form},
\cite{Vermaseren:2000nd,Larin:1991fz}, is devised to extract the
divergences from massless $2$-point functions. The propagators of
the massless fields in an arbitrary linear covariant gauge are
\begin{eqnarray}
\langle A^a_\mu(p) A^b_\nu(-p) \rangle &=& -~
\frac{\delta^{ab}}{p^2} \left[
\delta_{\mu\nu} ~-~ (1-\alpha) \frac{p_\mu p_\nu}{p^2} \right]\;, \nonumber \\
\langle c^a(p) {\overline c}^b(-p) \rangle &=&
\frac{\delta^{ab}}{p^2} ~~~,~~~
\langle \psi(p) {\overline{ \psi}}(-p) \rangle ~=~ \frac{\pslash}{p^2}\;, \nonumber \\
\langle B^a_{\mu\nu}(p) {\overline B}^b_{\sigma\rho}(-p) \rangle &=&
-~ \frac{\delta^{ab}}{2p^2} \left[ \delta_{\mu\sigma}
\delta_{\nu\rho} ~-~
\delta_{\mu\rho} \delta_{\nu\sigma} \right] \;,\nonumber \\
\langle G^a_{\mu\nu}(p) {\overline G}^b_{\sigma\rho}(-p) \rangle &=&
-~ \frac{\delta^{ab}}{2p^2} \left[ \delta_{\mu\sigma}
\delta_{\nu\rho} ~-~ \delta_{\mu\rho} \delta_{\nu\sigma} \right]\;,
\end{eqnarray}
where $p$ is the momentum. The necessary Feynman diagrams were
generated automatically with {\sc Qgraf} \cite{Nogueira:1991ex}.

We first checked that the same two loop anomalous dimensions emerge
for the gluon, Faddeev-Popov ghost and quarks in an arbitrary linear
covariant gauge as when the extra localizing fields are absent. It
was also explicitly verified that the correct coupling constant
renormalization constant is found. These results are in agreement
with the general argument of the previous subsection.

We have implemented the properties (\ref{jacobigen}) and
(\ref{abcd}) of the $\lambda^{abcd}$ coupling in a {\sc Form}
module, while it was assumed that
\begin{eqnarray}
\lambda^{acde} \lambda^{bcde} &=& \frac{1}{\NA} \delta^{ab}
\lambda^{pqrs} \lambda^{pqrs} ~~~,~~~ \nonumber\\\lambda^{acde}
\lambda^{bdce} &=& \frac{1}{\NA} \delta^{ab} \lambda^{pqrs}
\lambda^{prqs}\;,
\end{eqnarray}
which follows from the fact that there is only one rank 2 invariant
tensor in a classical Lie group.

At two loops in the $\MSbar$ scheme, we find that \begin{widetext}
\begin{eqnarray}
\gamma_B(a,\lambda) ~=~ \gamma_G(a,\lambda) &=& ( \alpha - 3 ) a ~+~
\left[ \left( \frac{\alpha^2}{4} + 2 \alpha - \frac{61}{6} \right)
C_A^2 ~+~ \frac{10}{3} T_F \Nf \right] a^2  +~ \frac{1}{128N_A}
\lambda^{abcd} \lambda^{acbd}\;, \label{gammab}
\end{eqnarray}
\end{widetext}
where $N_A$ is the dimension of the adjoint representation of the
colour group, $a=\frac{g^2}{16\pi^2}$ and we have also absorbed a
factor of $\frac{1}{4\pi}$ into $\lambda^{abcd}$ here and in later
anomalous dimensions. These anomalous dimensions are consistent with
the general observation that these fields must have the same
renormalization constants, in agreement with the output of the Ward
identities \cite{Capri:2005dy}. A check on (\ref{gammab}) is that
after the renormalization of the $3$-point gluon $B^a_{\mu\nu}$
vertex, the correct gauge parameter independent coupling constant
renormalization constant emerges.

We also determined the one loop $\beta$-function for the
$\lambda^{abcd}$ couplings. As this is present in a quartic
interaction it means that to deduce its renormalization constant, we
need to consider a $4$-point function. However, in such a situation
the {\sc Mincer} algorithm is not applicable since two external
momenta have to be nullified and this will lead to spurious infrared
infinities which could potentially corrupt the renormalization
constant. Therefore, for this renormalization only, we have resorted
to using a temporary mass regularization introduced into the
computation using the algorithm of \cite{Chetyrkin:1997fm} and
implemented in {\sc Form}. Consequently, we find the gauge parameter
independent anomalous dimension
\begin{widetext}
\begin{eqnarray}\label{betaabcd}
\beta^{abcd}_\lambda(a,\lambda) &=& \left[ \frac{1}{4} \left(
\lambda^{abpq} \lambda^{cpdq} + \lambda^{apbq} \lambda^{cdpq} +
\lambda^{apcq} \lambda^{bpdq}
+ \lambda^{apdq} \lambda^{bpcq} \right) \right. \nonumber \\
&& \left. -~ 12 C_A \lambda^{abcd} a ~+~ 8 C_A f^{abp} f^{cdp} a^2
~+~ 16 C_A f^{adp} f^{bcp} a^2 ~+~ 96 d_A^{abcd} a^2 \right]\;,
\end{eqnarray}
\end{widetext}
from both the $\lambda^{abcd} \overline{B}^a_{\mu\nu} B^{b\,\mu\nu}
\overline{B}^c_{\sigma\rho} B^{d\,\sigma\rho}$ and $\lambda^{abcd}
\overline{B}^a_{\mu\nu} B^{b\,\mu\nu} \overline{G}^c_{\sigma\rho}
G^{d\,\sigma\rho}$ vertices where $d_A^{abcd}$ is the totally
symmetric rank four tensor defined by
\begin{equation}
d_A^{abcd} ~=~ \mbox{Tr} \left( T^a_A T^{(b}_A T^c_A T^{d)}_A
\right)\;,
\end{equation}
with $T^a_A$ denoting the group generator in the adjoint
representation, \cite{vanRitbergen:1998pn}.  Producing the same
expression for both these $4$-point functions, aside from the gauge
independence, is a strong check on their correctness as well as the
correct implementation of the group theory. Moreover, as it should
be, $\beta^{abcd}$ enjoys the same symmetry properties as the tensor
$\lambda^{abcd}$, summarized in (\ref{abcd}).

We notice that $\lambda^{abcd}$~$=$~$0$ is not a fixed point due to
the extra $\lambda^{abcd}$-independent terms. If we had not included
the $\lambda^{abcd}$-interaction term in the original action, then
such a term would inevitably be generated at one loop through
quantum corrections, meaning that in this case there would have been
a breakdown of renormalizability.

Finally, we turn to the two loop renormalization of the mass $m$.
The corresponding operator can be read off from
(\ref{completeaction}) and is given by
\begin{equation}\label{nloa}
    \mathcal{M}=\left(B_{\mu\nu}^a-\oB_{\mu\nu}^a\right)F_{\mu\nu}^a\;.
\end{equation}
We insert $\mathcal{M}$ into a $A^a_\mu$-$B^b_{\nu\sigma}$ $2$-point
function and deduce the appropriate renormalization constant,
leading to the $\MSbar$ anomalous dimension
\begin{widetext}
\begin{eqnarray}\label{resgammaop}
\gamma_{\cal O}(a,\lambda) =
&-&2\left(\frac{2}{3}T_FN_f-\frac{11}{6}C_A\right)a
    -\left(\frac{4}{3}T_FN_fC_A+4T_FN_fC_F-\frac{77}{12}C_A^2\right)a^2
    +\frac{1}{8N_A}f^{abe}f^{cde}\lambda^{adbc}a-\frac{1}{128N_A}\lambda^{abcd}\lambda^{abcd}
\end{eqnarray}
\end{widetext}
as the two loop $\MSbar$ anomalous dimension. The gauge parameter
independence is again a good check, as the operator $\mathcal{M}$ is
gauge invariant.

\section{Conclusions}
We added a nonlocal mass term (\ref{massop}) to the Yang-Mills
action (\ref{1}), and starting from this, we succeeded in
constructing a renormalizable massive gauge model, which is gauge
invariant at the classical level and when quantized it enjoys a
nilpotent BRST symmetry. This BRST symmetry ensures that the
expectation value of gauge invariant operator is gauge parameter
independent. We have also proven the equivalence of the massless
version of our model with Yang-Mills gauge theories making use of a
``supersymmetry'' existing between the extra fields in that case. We
presented explicit two loop renormalization functions, thereby
verifying that the anomalous dimensions of the original Yang-Mills
quantities remain unchanged.

Many things could be investigated in the future concerning the gauge
model described by (\ref{completeaction}).

At the perturbative level, it could be investigated which
(asymptotic) states belong to a physical subspace of the model, and
in addition one should find out whether this physical subspace can
be endowed with a positive norm, which would imply unitarity. The
nilpotent BRST symmetry (\ref{brst3}) might be useful for this.

The model (\ref{completeaction}) is also asymptotically free,
implying that at low energies nonperturbative effects, such as
confinement, could set in. Proving and understanding the possible
confinement mechanism in this model is probably as difficult as for
usual Yang-Mills gauge theories.

It would also be interesting to find out whether a dynamically
generated term $m(\overline{B}-B)F$ might emerge, which in turn
could influence the gluon Green functions. This might also be
relevant in the context of gauge invariant $\frac{1}{Q^2}$ power
corrections, an issue that recently has also attracted attention
from the gauge/string duality side, the so-called AdS/QCD
\cite{Andreev:2006vy,Csaki:2006ji}.

\section*{Acknowledgments}
D.~Dudal and H.~Verschelde are grateful to the organizers of the
stimulating \emph{IRQCD06} conference. The Conselho Nacional de
Desenvolvimento Cient\'{i}fico e Tecnol\'{o}gico (CNPq-Brazil), the
Faperj, Funda{\c{c}}{\~{a}}o de Amparo {\`{a}} Pesquisa do Estado do
Rio de Janeiro, the SR2-UERJ and the Coordena{\c{c}}{\~{a}}o de
Aperfei{\c{c}}oamento de Pessoal de N{\'\i}vel Superior (CAPES) are
gratefully acknowledged for financial support.


\begin{thebibliography}{99}
\bibitem{Gross:1973id}
D.~J.~Gross and F.~Wilczek, Phys.\ Rev.\ Lett.\ {\bf 30} (1973)
1343.

\bibitem{Politzer:1973fx}
H.~D.~Politzer, Phys.\ Rev.\ Lett.\ {\bf 30} (1973) 1346.

\bibitem{Shifman:1978bx}
M.~A.~Shifman, A.~I.~Vainshtein and V.~I.~Zakharov, Nucl.\ Phys.\ B
{\bf 147} (1979) 385.

\bibitem{Gubarev:2000eu}  F.~V.~Gubarev, L.~Stodolsky and V.~I.~Zakharov,
Phys.\ Rev.\ Lett.\ \textbf{86} (2001) 2220.

\bibitem{Gubarev:2000nz}  F.~V.~Gubarev and V.~I.~Zakharov,
Phys.\ Lett.\ B \textbf{501} (2001) 28.

\bibitem{Lavelle:1995ty}
M.~Lavelle and D.~McMullan, Phys.\ Rept.\  {\bf 279} (1997) 1.

\bibitem{Capri:2005dy}
M.~A.~L.~Capri, D.~Dudal, J.~A.~Gracey, V.~E.~R.~Lemes,
R.~F.~Sobreiro, S.~P.~Sorella and H.~Verschelde, Phys.\ Rev.\ D {\bf
72} (2005) 105016.

\bibitem{Gribov:1977wm}
V.~N.~Gribov, Nucl.\ Phys.\ B {\bf 139} (1978) 1.

\bibitem{Semenov}  Semenov-Tyan-Shanskii and V.A. Franke, Zapiski Nauchnykh
Seminarov Leningradskogo Otdeleniya Matematicheskogo Instituta im.
V.A. Steklov AN SSSR, Vol. \textbf{120} (1982) 159. English
translation: New York: Plenum Press 1986.

\bibitem{Zwanziger:1990tn}  D.~Zwanziger,
Nucl.\ Phys.\ B \textbf{345}, 461 (1990).

\bibitem{Dell'Antonio:1989jn}  G.~Dell'Antonio and D.~Zwanziger,
Nucl.\ Phys.\ B \textbf{326}, 333 (1989).

\bibitem{Dell'Antonio:1991xt}  G.~Dell'Antonio and D.~Zwanziger,
Commun.\ Math.\ Phys.\ \textbf{138}, 291 (1991).

\bibitem{vanBaal:1991zw}  P.~van Baal,
Nucl.\ Phys.\ B \textbf{369}, 259 (1992).

\bibitem{Dudal:2002pq}  D.~Dudal, H.~Verschelde and S.~P.~Sorella,
Phys.\ Lett.\ B \textbf{555} (2003) 126.

\bibitem{Gracey:2002yt}
J.~A.~Gracey, Phys.\ Lett.\ B {\bf 552} (2003) 101.

\bibitem{Verschelde:2001ia}  H.~Verschelde, K.~Knecht, K.~Van Acoleyen and
M.~Vanderkelen, Phys.\ Lett.\ B \textbf{516} (2001) 307.

\bibitem{Browne:2004mk}  R.~E.~Browne and J.~A.~Gracey,
Phys.\ Lett.\ B \textbf{597} (2004) 368.

\bibitem{Gracey:2004bk}  J.~A.~Gracey,
Eur.\ Phys.\ J.\ C \textbf{39} (2005) 61.

\bibitem{Browne:2003uv}  R.~E.~Browne and J.~A.~Gracey,
JHEP \textbf{0311} (2003)  029.

\bibitem{Lavelle:1988eg}  M.~J.~Lavelle and M.~Schaden, Phys.\ Lett.\ B \textbf{208} (1988) 297.

\bibitem{Kondo:2001nq}  K.~I.~Kondo,
Phys.\ Lett.\ B \textbf{514} (2001) 335.

\bibitem{Boucaud:2001st}  P.~Boucaud, A.~Le Yaouanc, J.~P.~Leroy,
J.~Micheli, O.~Pene and J.~Rodriguez-Quintero, Phys.\ Rev.\ D
\textbf{63} (2001) 114003.

\bibitem{Boucaud:2002nc}  P.~Boucaud, J.~P.~Leroy, A.~Le Yaouanc, J.~Micheli, O.~Pene, F.~De Soto, A.~Donini, H.~Moutard and
J.~Rodriguez-Quintero, Phys.\ Rev.\ D \textbf{66} (2002) 034504.

\bibitem{RuizArriola:2004en}  E.~Ruiz Arriola, P.~O.~Bowman and
W.~Broniowski, Phys.\ Rev.\ D \textbf{70} (2004) 097505.

\bibitem{Furui:2005bu}  S.~Furui and H.~Nakajima, hep-lat/0503029.

\bibitem{Furui:12ks}
S.~Furui and H.~Nakajima, Phys.\ Rev.\ D {\bf 73} (2006) 074503.

\bibitem{Marenzoni:1994ap} P.~Marenzoni, G.~Martinelli and N.~Stella,
Nucl.\ Phys.\ B {\bf 455} (1995) 339.

\bibitem{Leinweber:1998uu}
D.~B.~Leinweber, J.~I.~Skullerud, A.~G.~Williams and C.~Parrinello
[UKQCD Collaboration],  Phys.\ Rev.\ D {\bf 60} (1999) 094507
  [Erratum-ibid.\ D {\bf 61} (2000) 079901].

\bibitem{Langfeld:2001cz}
K.~Langfeld, H.~Reinhardt and J.~Gattnar, Nucl.\ Phys.\ B {\bf 621}
(2002) 131.

\bibitem{Amemiya:1998jz}
K.~Amemiya and H.~Suganuma, Phys.\ Rev.\ D {\bf 60} (1999) 114509.

\bibitem{Bornyakov:2003ee}
V.~G.~Bornyakov, M.~N.~Chernodub, F.~V.~Gubarev, S.~M.~Morozov and
M.~I.~Polikarpov, Phys.\ Lett.\ B {\bf 559} (2003) 214.

\bibitem{Cucchieri:2004mf}
A.~Cucchieri, T.~Mendes and A.~R.~Taurines, Phys.\ Rev.\ D {\bf 71}
(2005) 051902.

\bibitem{Parisi:1980jy}
G.~Parisi and R.~Petronzio, Phys.\ Lett.\ B {\bf 94} (1980) 51.

\bibitem{Field:2001iu}
J.~H.~Field, Phys.\ Rev.\ D {\bf 66} (2002) 013013.

\bibitem{Ruegg:2003ps}
H.~Ruegg and M.~Ruiz-Altaba, Int.\ J.\ Mod.\ Phys.\ A {\bf 19}
(2004) 3265.

\bibitem{vanDam:1970vg}
H.~van Dam and M.~J.~G.~Veltman, Nucl.\ Phys.\ B {\bf 22} (1970)
397.

\bibitem{Sobreiro:2004us}  R.~F.~Sobreiro, S.~P.~Sorella, D.~Dudal and
H.~Verschelde, Phys.\ Lett.\ B \textbf{590} (2004) 265.

\bibitem{Dudal:2005na}  D.~Dudal, R.~F.~Sobreiro, S.~P.~Sorella and
H.~Verschelde, Phys.\ Rev.\ D \textbf{72} (2005) 014016.

\bibitem{Dudal:2004rx}  D.~Dudal, J.~A.~Gracey, V.~E.~R.~Lemes,
M.~S.~Sarandy, R.~F.~Sobreiro, S.~P.~Sorella and H.~Verschelde,
Phys.\ Rev.\ D \textbf{70} (2004) 114038.

\bibitem{Capri:2005tj}  M.~A.~L.~Capri, V.~E.~R.~Lemes, R.~F.~Sobreiro,
S.~P.~Sorella and R.~Thibes, Phys.\ Rev.\ D {\bf 72} (2005) 085021.

\bibitem{Dudal:2003np}  D.~Dudal, H.~Verschelde, V.~E.~R.~Lemes,
M.~S.~Sarandy, R.~F.~Sobreiro, S.~P.~Sorella and J.~A.~Gracey,
Phys.\ Lett.\ B \textbf{574} (2003) 325.

\bibitem{Dudal:2003by}  D.~Dudal, H.~Verschelde, J.~A.~Gracey,
V.~E.~R.~Lemes, M.~S.~Sarandy, R.~F.~Sobreiro and S.~P.~Sorella,
JHEP \textbf{0401} (2004) 044.

\bibitem{Sobreiro:2005vn}  R.~F.~Sobreiro and S.~P.~Sorella, JHEP \textbf{0506}
(2005) 054.

\bibitem{Dudal:2003gu}  D.~Dudal, H.~Verschelde, V.~E.~R.~Lemes,
M.~S.~Sarandy, S.~P.~Sorella and M.~Picariello, Annals Phys.\
\textbf{308} (2003) 62.

\bibitem{Dudal:2003pe}  D.~Dudal, H.~Verschelde, V.~E.~R.~Lemes, M.~S.~Sarandy, R.~F.~Sobreiro,
S.~P.~Sorella, M.~Picariello and J.~A.~Gracey, Phys.\ Lett.\ B
\textbf{569} (2003) 57.

\bibitem{Jackiw:1997jg}
R.~Jackiw and S.~Y.~Pi, Phys.\ Lett.\ B {\bf 403} (1997) 297.

\bibitem{Zwanziger:1989mf}
D.~Zwanziger, Nucl.\ Phys.\ B \textbf{323} (1989) 513.

\bibitem{Zwanziger:1992qr}
D.~Zwanziger, Nucl.\ Phys.\ B \textbf{399} (1993) 477.

\bibitem{Piguet:1995er}
O.~Piguet and S.~P.~Sorella, Lect.\ Notes Phys.\ \textbf{M28} (1995)
1.

\bibitem{Capri:2006ne}
M.~A.~L.~Capri, D.~Dudal, J.~A.~Gracey, V.~E.~R.~Lemes,
R.~F.~Sobreiro, S.~P.~Sorella and H.~Verschelde, Phys.\ Rev.\ D {\bf
74} (2006) 045008.

\bibitem{Gorishnii:1989gt}
 S.~G.~Gorishnii, S.~A.~Larin, L.~R.~Surguladze and F.~V.~Tkachov, Comput.\ Phys.\ Commun.\  {\bf 55} (1989) 381.

\bibitem{Vermaseren:2000nd}
J.~A.~M.~Vermaseren, math-ph/0010025.

\bibitem{Larin:1991fz}
S.~A.~Larin, F.~V.~Tkachov and J.~A.~M.~Vermaseren, \emph{The FORM
version of MINCER}, NIKHEF-H-91-18.

\bibitem{Nogueira:1991ex}
P.~Nogueira, J.\ Comput.\ Phys.\  {\bf 105} (1993) 279.

\bibitem{Chetyrkin:1997fm}
K.~G.~Chetyrkin, M.~Misiak and M.~Munz, Nucl.\ Phys.\ B {\bf 518}
(1998) 473.

\bibitem{vanRitbergen:1998pn}
T.~van Ritbergen, A.~N.~Schellekens and J.~A.~M.~Vermaseren, Int.\
J.\ Mod.\ Phys.\ A {\bf 14} (1999) 41.

\bibitem{Andreev:2006vy}
O.~Andreev, Phys.\ Rev.\ D {\bf 73} (2006) 107901.

\bibitem{Csaki:2006ji}
C.~Csaki and M.~Reece,  hep-ph/0608266.


\end{thebibliography}
\end{document}